\documentclass[a4paper, 10pt, twocolumn]{article}

\usepackage{amsfonts,amssymb,amsmath,amsthm}
\usepackage{subfigure}
\usepackage{graphicx}
\usepackage[footnotesize]{caption}

\usepackage[top=1.5cm, left=1.5cm, right=1.5cm, bottom=1.5cm]{geometry}

\title{MIP and Set Covering approaches for Sparse Approximation\footnote{This research was partially supported by Labex DigiCosme (project ANR-11-LABEX-0045-DIGICOSME) operated by ANR as part of the program ``Investissement d'Avenir'' Idex Paris-Saclay (ANR-11-IDEX-0003-02).}}

\author{
Diego Delle Donne$^1$, Matthieu Kowalski$^2$ and Leo Liberti$^1$.\\
\footnotesize $^1$LIX CNRS, Ecole Polytechnique, Institut Polytechnique de Paris, 91128 Palaiseau, France.\\
\footnotesize $^2$ Laboratoire des Signaux et Systèmes, UMR 8506 Univ Paris-Sud -- CNRS -- Centralesupelec, 91192
Gif-sur-Yvette Cedex, France
}
\date{\empty} 

\renewenvironment{abstract}{\bf\small {\em\ Abstract---}}{}

\newcommand{\R}{\mathbb{R}}

\newtheorem{theorem}{Theorem}[section]

\newtheorem{proposition}[theorem]{Proposition}

\begin{document}

\maketitle

\begin{abstract} The \emph{sparse approximation} problem asks to find a solution $x$ such that $||y - Hx|| < \alpha$, for a given norm $||\cdot||$, minimizing the size of the support $||x||_0 := \#\{j \ |\ x_j \neq 0 \}$.
	We present valid inequalities for Mixed Integer Programming (MIP) formulations for this problem and we show that these families are sufficient to describe the set of feasible supports. 
	This leads to a reformulation of the problem as an Integer Programming (IP) model which in turn represent a \emph{minimum set covering} formulation, thus yielding many families of valid inequalities which may be used to strengthen the models up.
	We propose algorithms to solve sparse approximation problems including a branch \& cut for the MIP, a two-stages algorithm to tackle the set covering IP and a heuristic approach based on Local Branching type constraints. 
	These methods are compared in a computational experimentation with the goal of testing their practical potential.
\end{abstract}

\section{Introduction}
The \emph{sparse representation} of a vector $y \in \R^n$ in a \emph{dictionary} $H \in \R^{n\times m}$ aims to find a solution $x \in \R^m$ to the system $Hx = y$, having the minimum number of non-zero components, i.e., minimizing the so-called $\ell_0$ pseudo-norm of $x$, defined by $||x||_0 := |\{j\;|\;x_j\not=0\}|$.
The \emph{sparse approximation} problem takes also into account \emph{noise} and model errors. It relaxes the equality constraint aiming to minimize the misfit data measure $||y - Hx||$, for a given norm $||\cdot||$. In this context, several optimization problems may be stated as such:
\begin{enumerate}
	\item minimize $||x||_0$ subject to a given \emph{threshold} for the data misfit $||y - Hx|| \leq \alpha$,
	\item minimize the data misfit $||y - Hx||$ subject to a given bound $||x||_0 \leq k$,
	\item minimize a weighted sum $\lambda_1 ||y - Hx|| + \lambda_2 ||x||_0$ for some $\lambda_1, \lambda_2 \in \R$.
\end{enumerate}

\noindent
In this work, we study mixed integer programming (MIP) formulations for the problem stated in Item 1 when the norm used for the data misfit measure is the  $\ell_1$ and $\ell_\infty$ norms (we address the reader to \cite{bourguignon:hal-01254856} for the rest of the cases). 
Following the notation from \cite{bourguignon:hal-01254856}, we define these problems as
$$
\mathcal P_{0/p}(\alpha) : \min_{x} ||x||_0 \ \text{ s. t. } ||y - H x||_p \leq \alpha,
$$
In the remaining, we may write just $\mathcal P_{0/p}$ whenever $\alpha_p$ is clear from the context and/or irrelevant. 
Also, for any natural number $t$, we may use $[t]$ as a shortcut for the set $\{1, \dots, t\}$.\\

Some natural \emph{mixed-integer programming} (MIP) formulations for $\mathcal P_{0/1}$ and $\mathcal P_{0/\infty}$ are introduced in \cite{bourguignon:hal-01254856}. These models use decision variables $x_j \in \R$, for each $j\in [m]$ to determine the solution and binary \emph{support} variables $b_j$ to state whether $x_j$ has a non-zero value or not.
They require to (artificially) bound $|x_j|$ with a value $M$ in order to properly state the models.
Then, $\sum_{j\in [m]} b_j$ is minimized subject to appropriate constraints.
We call these formulations $\mathsf{MIP}_{0/1}$ and $\mathsf{MIP}_{0/\infty}$, respectively, and we state them here for completeness.
\\ [-0.4cm]
\begin{align}
[\mathsf{MIP}_{0/1}]\qquad  \min \sum_{j\in [m]} & b_j \label{mod:p01:obj} \\
-M b_j \leq x_j & \leq M b_j &  \forall j \in [m] \label{mod:p01:1}  \\
-w_i \leq y_i - \sum_{j \in [m]} h_{ij} x_j & \leq w_i &  \forall i \in [n] \label{mod:p01:2}  \\
\sum_{i \in [n]} w_i & \leq \alpha_1  &  \label{mod:p01:3}  \\
w_i, x_j \in \R, \ b_j & \in \{0,1\}  &  \forall i \in [n], j \in [m] \label{mod:p01:4} 
\end{align}
\\ [-0.8cm]
\begin{align}
[\mathsf{MIP}_{0/\infty}]\qquad  \min \sum_{j\in [m]} & b_j \label{mod:p0inf:obj} \\
-M b_j \leq x_j & \leq M b_j &  \forall j \in [m] \label{mod:p0inf:1}  \\
-w \leq y_i - \sum_{j \in [m]} h_{ij} x_j & \leq w &  \forall i \in [n] \label{mod:p0inf:2}  \\
w & \leq \alpha_{\infty} &  \label{mod:p0inf:3}  \\
x_j \in \R, \ b_j & \in \{0,1\}  &  \forall j \in [m] \label{mod:p0inf:4} 
\end{align}
In \cite{bourguignon:hal-01254856}, these formulations are solved directly by CPLEX.
As far as we know, no polyhedral studies have been done for these formulations, with the goal of developing more powerful resolution (e.g., cutting planes based) algorithms.

In this context, we study the polytopes arising from these formulations and derive valid inequalities for them which we then use within an initial branch and cut algorithm for $\mathsf{MIP}_{0/p}$.
Additionally, we prove that the obtained inequalities are sufficient to describe the projection of these polytopes into the space of the binary variables $b_j$ and that this projections are in fact \emph{set covering} polytopes.
Based on this fact, we introduce a novel IP approach for $\mathcal P_{0/p}$ which consists in solving a pure combinatorial set covering formulation (with exponentially many covering constraints) and we propose a two-stages algorithm to tackle this IP.
Furthermore, we propose a heuristic approach resorting to the Variable Neighborhood Search metaheuristic \cite{vns} and Local Branching \cite{localbranching} type constraints. 

\section{Valid inequalities and a Set Covering formulation}

We say that a set of columns $J \subseteq [m]$ is a \emph{forbidden support} for $\mathcal P_{0/p}$ if there exist no solutions with $J$ as support, i.e., if $\min_x \{ ||y - H^Jx^J||_p\} > \alpha_p$, where $H^J$ (resp. $x^J$) is the submatrix of $H$ (resp. subvector of $x$) involving only those columns indexed by $J$.


\begin{proposition}\label{prop:forbiddensupineq}
	If $J \subseteq [m]$ is a forbidden support for $\mathcal P_{0/p}$, then the \emph{forbidden support inequality}
	\begin{equation} \label{eq:forbiddensup}
	\sum_{j \in [m] \setminus J} b_j \geq 1	
	\end{equation}
	is valid for $MIP_{0/p}$.	
\end{proposition}


\noindent
From Proposition \ref{prop:forbiddensupineq}, we derive some subfamilies of valid inequalities for which we developed separation procedures (both exact and heuristics) and implemented a \emph{branch \& cut} algorithm using them as cutting planes.
We omit here these elements due to space limitations.
We state next an interesting theoretical result about the \emph{forbidden support inequalities} \eqref{eq:forbiddensup}.

\begin{proposition}\label{prop:equiv_formulations}
	The projection on the variables $b_j$ of all feasible solutions of formulation $MIP_{0/p}$ can be described by the forbidden support inequalities \eqref{eq:forbiddensup} as 
	\begin{align*}
	\mathcal{P}_{fs} = \{ b &\in \{0,1\}^m \ |\\ & b \text{ satisfies \eqref{eq:forbiddensup} for each forb. supp. $J \subseteq [m]$} \}.
	\end{align*}
\end{proposition}


\noindent
Proposition \ref{prop:equiv_formulations} lets us obtain a minimum support $\hat b$ of a solution to $\mathcal{P}_{0/p}$ by solving the following \emph{integer programming} (IP) formulation:
\begin{align}
	[IP^{cov}_{0/p}]\qquad  &\min \sum_{j\in [m]} b_j \label{mod:p0p_cov:obj} \\
	\sum_{j \in [m] \setminus J} b_j &\geq 1	& \forall \text{ forb. supp. } J \subseteq [m] \label{mod:p0p_cov:1} \\
	b_j & \in \{0,1\}  &  \forall j \in [m] \label{mod:p0p_cov:2} 
\end{align}

\noindent
We remark that by solving $\mathsf{IP}^{cov}_{0/p}$ we do not obtain a solution for $\mathcal{P}_{0/p}$ but just an optimal support $S \subseteq [m]$. However, a solution $x$ for this support can be efficiently obtained afterwards. 
Precisely, the non-zero values of $x$ can be obtained by minimizing $||y - H^Sx^S||_p$ (which for $p\in \{1, \infty\}$ can be achieved by solving a linear program).
Moreover, as the support is already fixed for this last step, there is no need to use the (usually artificial) big-M bounds for $x$, which gives an important advantage against formulation $\mathsf{MIP}_{0/p}$.
\\

An initial drawback towards the computational resolution of $\mathsf{IP}^{cov}_{0/p}$, is that the formulation may have exponentially many constraints \eqref{mod:p0p_cov:1}. 
However, we note that we can efficiently test whether a vector $b \in \{0,1\}^m$ satisfies all these constraints or not, without the need of enumerating them. 
Due to Proposition \ref{prop:equiv_formulations}, a vector $b \in \{0,1\}^m$ satisfies \eqref{mod:p0p_cov:1} if and only if there exists a feasible solution to $\mathsf{MIP}_{0/p}$ with support $b$, i.e., 
$\min_{x \in \R^m} \{ ||y - H^Sx^S||_p\} \leq \alpha_p$, where $S$ is the support described by $b$. As mentioned before, for $p\in \{1, \infty\}$, this can be tested by solving a linear program.
Based on this characteristic, we propose a two stages algorithm which starts from a \emph{combinatorial relaxation} of $\mathsf{IP}^{cov}_{0/p}$ with a few (or none) constraints and dynamically adds constraints \eqref{mod:p0p_cov:1} whenever an optimal integer but not feasible solution is found.\\

The equivalence given by Proposition \ref{prop:equiv_formulations} has also useful implications. In particular, we note that $\mathsf{IP}^{cov}_{0/p}$ represents a \emph{minimum set covering} problem and this kind of problems has been widely studied in the literature both in the polyhedral and in the combinatorial aspects \cite{Balas89,Balas89b,Borndoerfer1998,Cornuejols89,Laurent89,Nobili89,SanchezGarcia98,Sassano89}.
A direct implication of this is the fact that any valid inequality for these \emph{set covering polytopes} can provide a valid inequality for $\mathsf{MIP}_{0/p}$. We give next an example in which we depict a (non intuitive) family of valid inequalities for $\mathsf{MIP}_{0/p}$ obtained from a known family of set covering facets \cite{Balas89}.

\begin{proposition}\label{prop:forbiddensupfamineq}
	Let $\mathcal{J} \subseteq 2^{[m]}$ be a family of forbidden supports for $\mathcal P_{0/p}$, and define
	$J^{none} := [m] \setminus \bigcup_{J \in \mathcal{J}} J$, and 
	$J^{some} := [m] \setminus (J^{none} \cup \bigcap_{J \in \mathcal{J}} J)$.
	Then the \emph{forbidden support family inequality}
	\begin{equation} \label{eq:forbiddensupfam}
	\sum_{j \in J^{none}} 2 b_j + \sum_{j \in J^{some}} b_j \geq 2	
	\end{equation}
	is valid for $\mathsf{MIP}_{0/p}$.	
\end{proposition}

\section{Local Branching based heuristic}\label{sec:vns}

Given an integer solution $(\hat x, \hat w, \hat b)$ for $\mathsf{MIP}_{0/p}$, the idea of \emph{local branching} is to impose a constraint forcing the solution to be ``similar'' to $(\hat x, \hat w, \hat b)$. In our setting, this constraint ensures that the difference in the support sizes should not exceed a prespecified value $\delta$, i.e., 
\begin{equation} \label{eq:localbranching}
\sum_{j \in J_{0}} b_j + \sum_{j \in J_{1}} (1 - b_j) \leq \delta.	
\end{equation}
with $J_i := \{ j \in [m] : \hat{b}_j = i \}$, for $i \in \{0,1\}$.
%
%
The addition of Constraint \eqref{eq:localbranching} to $\mathsf{MIP}_{0/p}$ reduces the feasible region to a sort of \emph{$\delta$-neighborhood} of the given point $(\hat x, \hat w, \hat b)$, aiming to obtain a faster (although heuristic) resolution, which should be the case for small values of $\delta$.
We propose a heuristic algorithm (based on the Variable Neighborhood Search \cite{vns} metaheuristic) which starts from an initial solution and explores its \emph{$\delta$-neighborhood} for increasing values of $\delta$, beginning from a predefined value $\delta_0$. Every time a better solution is found, $\delta$ is reseted to $\delta_0$ and the process is repeated from this new solution. 

\section{Final remarks}
We proposed some MIP based approaches for the sparse approximation problem, including both exact and heuristic methods. 
All these methods can be improved and/or combined towards the development of a general algorithm.
As a first step on that direction, in this work we aimed to test the potential of the proposed methods by conducting a computational experimentation over a set of hard instances (arising from a pathological example from the literature). 
We omit to show numerical results here due to space limitations.
According to our experimentation, the heuristic algorithm (from Section \ref{sec:vns}) provides a good starting point towards the development of an efficient approximation algorithm for the problem addressed in this work.	

Within this work, an interesting relation between $\mathcal{P}_{0/p}$ and the minimum set covering problem was established. As a future line of research, we believe that this relation shall be exploited. In particular, it would be interesting to derive more families of valid inequalities from set covering polytopes, in order to strengthen the initial branch and cut implemented for this work. Additionally, the combinatorial aspects of the minimum set covering problem may uncover interesting tools towards the efficient resolution of $\mathsf{MIP}_{0/p}$. We leave these aspects for a future work.


\bibliography{biblio}
\bibliographystyle{plain}

\end{document}